\documentclass[12pt,a4paper]{conference}

\usepackage{fancyhdr}
\usepackage{graphicx,amsmath,amssymb,cite}
\usepackage{multind}
\makeindex{author} \makeindex{subject}

\bigskip

\newcommand{\beq}{\begin{equation}}
\newcommand{\eeq}[1]{\label{#1}\end{equation}}
\newcommand{\eeqn}{\end{equation}}

\newcommand{\beqa}{\begin{eqnarray}}
\newcommand{\eeqa}[1]{\label{#1}\end{eqnarray}}
\newcommand{\eeqan}{\end{eqnarray}}

\let\bar=\overbar

\newcommand{\Dslash}{\not{\hbox{\kern-4pt $D$}}}
\newcommand{\dslash}{\not{\hbox{\kern-2pt $\del$}}}

\newcommand{\msb}{{\bar{\ssstyle M \kern -1pt S}}}

\begin{document}

\Chapter{QCD ANALYSIS FOR NUCLEAR PARTON DISTRIBUTIONS IN THE NEXT
TO LEADING ORDER}
           {QCD analysis for nuclear parton ...}{S. Atashbar Tehrani \it{et al.}} \vspace{-6
cm}\includegraphics[width=6 cm]{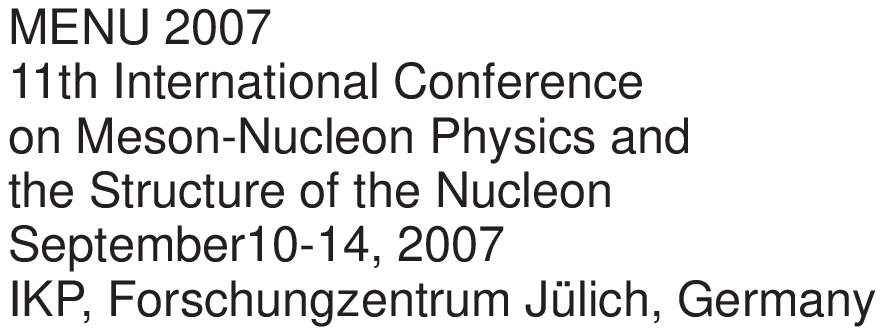}
\vspace{4 cm}

\addcontentsline{toc}{chapter}{{\it S. Atashbar Tehrani}}
\label{authorStart}

\begin{raggedright}

{\it S. Atashbar Tehrani}\index{author}{S. Atashbar Tehrani}\\
Physics Department, Semnan University, Semnan, Iran;\\
Institute for Studies in Theoretical Physics and Mathematics
(IPM), P.O.Box 19395-5531,
Tehran, Iran\\

{\it Ali N. Khorramian}\index{author}{Ali N. Khorramian}\\
Physics Department, Semnan University, Semnan, Iran; \\ Institute
for Studies in Theoretical Physics and Mathematics (IPM), P.O.Box
19395-5531,
Tehran, Iran\\

\end{raggedright}
\begin{center}
\textbf{Abstract}
\end{center}
A QCD analysis of the nuclear parton distributions and structure
functions in the NLO is performed by using the world data. By
having bounded parton distributions for a nuclear with atomic
number A, we can obtain the nuclear structure function in x space.
Our results for nuclear structure function ratio $F^A_2 /F^D_2$
for some different values of A, are in good agreement with the
experimental data. We compare our results for LO and NLO
approximation.
\section{Introduction}
Parametrization of nuclear parton distributions is investigated in
the next--to--leading order (NLO) of $\alpha_s$. Unpolarized
parton distributions in the nucleon are now well determined in the
region from very small $x$ to large $x$ by using various
experimental data. Initial distributions are assumed at a fixed
$Q^2$ with parameters which are determined by a $\chi^2$ analysis.
In this work we used the MRST parametrization~\cite{Martin:2006qz}
as the input parton distributions in the nucleon. In Ref.
\cite{Hirai:2001np} a LO QCD analysis was performed and authors
applied the MRST parton distributions~\cite{Martin:1999ww} in the
nucleon. Until now much efforts have been done to compute the
nuclear parton densities and structure functions in the
perturbative
QCD~\cite{Hirai:2001np,Tehrani:2004hp,Tehrani:2006gy,Eskola:1998df,Eskola:1998iy,
deFlorian:2003qf}. In this paper after parametrization of nuclear
parton distributions in $Q_0^2=4$ GeV$^2$, we will obtain the
nuclear structure function ratio $F^A_2 /F^D_2$ for helium, carbon
and calcium nuclei in the LO and NLO.
\section{Nuclear Structure Function}
Our analysis is done in the next--to--leading order of $\alpha_s$.
According to parton model the nuclear structure function $F_2^A$
in the NLO is given by~\cite{Gluck:1989ze}
\begin{eqnarray}
\frac{1}{x}F_{2}^{eA}(x,Q^{2})&&=\sum_{q=u,d,s}e_{q}^{2}\left\{ q^A(x,Q^{2})+%
\overline{q}^A(x,Q^{2})+\frac{\alpha _{s}(Q^{2})}{2\pi
}\right.\nonumber\\&& \left. \times \left[ c_{q,2}\otimes
(q^A+\overline{q}^A)+2c_{g,2}\otimes g^A\right]
\right\}~,\label{stfunction}
\end{eqnarray}
where $e_q$ is the quark charge, and $q^A({\bar q}^A)$ is the
quark (antiquark) distribution in the nucleus $A$. Here the sum
expands over all light quarks $q=u,~d,~s$ and $c_{q,2}$, $c_{g,2}$
are as following

\begin{eqnarray}
c_{q,2} &=&\frac{4}{3}\left[ \frac{1+z^{2}}{1-z}\left( \ln \frac{1-z}{z}-%
\frac{3}{4}\right) +\frac{1}{4}(9+5z)\right] _{+} \nonumber\\
c_{g,2} &=&\frac{1}{2}\left[\left( z^{2}+(1-z)^{2}\right) \ln \frac{1-z}{z}%
-1+8z(1-z)\right]~.\label{wilsoncof}
\end{eqnarray}
The convolutions are defined as
\begin{equation}
c\otimes q=\int_{x}^{1}\frac{dy}{y}c\left( \frac{x}{y}\right)
q(y,Q^{2})~.\label{convoloution}
\end{equation}
Notice that
\begin{equation}
\int_{x}^{1}\frac{dy}{y}f\left( \frac{x}{y}\right) _{+}g(y)=\int_{x}^{1}%
\frac{dy}{y}f\left( \frac{x}{y}\right) \left[
g(y)-\frac{x}{y}g(x)\right]
-g(x)\int_{0}^{x}dyf(y)~.\label{convoloutionplus}
\end{equation}
In this paper we assumed the flavor symmetric antiquark
distribution, ${\bar u}^A={\bar d}^A={\bar s}^A\equiv{\bar q}^A$.
We consider also the nuclear parton distributions as

\begin{eqnarray}
u_{v}^{A}={\cal
W}_{u_v}\frac{Z\;u_{v}+N\;d_{v}}{A},\;\;\;d_{v}^{A} ={\cal
W}_{d_v}\frac{Z\;d_{v}+N\;u_{v}}{A},\;\;\;\overline{q}^{A} ={\cal
W}_{s}\;\overline{q},\;\;\;g^{A} ={\cal W}_{g}\;g,
\label{nuclearparton}
\end{eqnarray}
in the above equations, we suppose the functional form for the
weight function for all partons as

\begin{equation}
{\cal
W}_{i} =1+\left( 1-\frac{1}{A^{1/3}}\right) \left( \frac{%
a_{i}+b_i~x+c_i x^{2}+d_i~x^{3}}{(1-x)^{e_i}}\right)~.
\label{wfunction}
\end{equation}

After using the MRST parton distributions in the nucleon at
$Q_0^2$=4 GeV$^2$ we can be able to determine some unknown
parameters which appear in the weight functions  by a $\chi^2$
analysis of the data on ration of nuclear structure
functions~\cite{Ashman:1988bf,Arneodo:1989sy,Gomez:1993ri,Amaudruz:1995tq,Arneodo:1995cs,Adams:1995is}.
In Fig. 1 we plot our QCD results for weight functions in the NLO,
which defined in Eqs. (\ref{nuclearparton}, \ref{wfunction}) for
deuteron, helium, carbon and calcium nuclei.
\begin{figure}
\begin{center}
\includegraphics[width=9 cm]{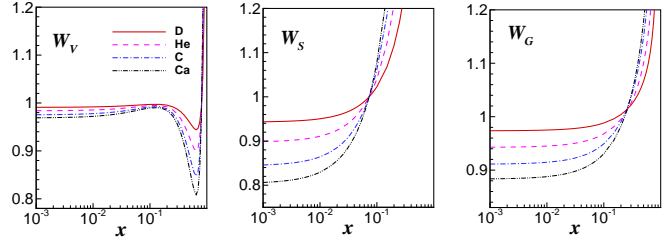}
\caption{QCD results for weight functions in the NLO for deuteron,
helium, carbon and calcium nuclei} \label{Fig:w}
\end{center}
\end{figure}
In Fig. 2 our results are compared with the experimental data at
$Q^2=5$ GeV$^2$ for helium, carbon and calcium nuclei in the LO
and NLO.
\begin{figure}
\begin{center}
\includegraphics[width=7.5 cm]{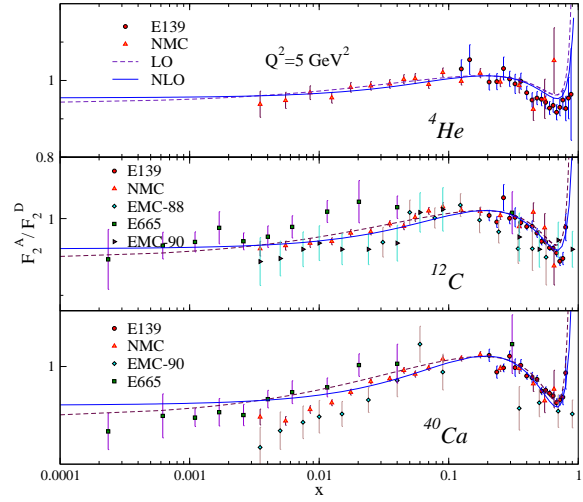}
\caption{Our QCD results are compared with the experimental data
at $Q^2=5$ GeV$^2$ for helium, carbon and calcium nuclei in the LO
and NLO.} \label{Fig:F2}
\end{center}
\end{figure}
\section*{Acknowledgments}
S.A.T. acknowledge the Persian Gulf university for the financial
support of this project.


 \end{document}